\documentclass[12pt]{article}
\usepackage{amssymb}
\usepackage{graphicx,psfrag}

\title{Finite Size Formula in the XYZ Spin Chain}
\author{Yasuhiro Fujii\footnotemark[2]\ \ and Miki Wadati\footnotemark[3]}
\date{\textit{Department of Physics, Graduate School of Science, \\
  University of Tokyo,\\  Hongo 7-3-1, Bunkyo-ku, Tokyo 113-0033, Japan}}

\begin{document}

\maketitle
\baselineskip 5.3mm

\footnotetext[2]{e-mail: \texttt{fujii@monet.phys.s.u-tokyo.ac.jp}}
\footnotetext[3]{e-mail: \texttt{wadati@monet.phys.s.u-tokyo.ac.jp}}

\begin{abstract}
  The XYZ spin chain with boundaries is studied.
  We construct the vacuum state by the vertex operators
  in the level one modules of the elliptic algebra,
  and compact it through a geometric symmetry of the model
  called the turning symmetry.
  From this simplified expression
  the ``finite size formula'' for magnetizations
  in the bounded and in the half-infinite chains is deduced.
  Applying this formula we calculate the spontaneous magnetization
  in the bounded XYZ model.
\end{abstract}

\section{Introduction}
The one-dimensional Heisenberg spin chains are
exactly solved with the aid of the Yang-Baxter equations
and are classified by the genus of those solutions.
For instance, the XXZ spin chain is realized
by its trigonometric (genus-$0$) solution,
which is controlled by quantum affine algebra $U_q(\widehat{sl}_2)$.
Its representation theory further yields
the idea of the vertex operator (VO),
which enables us to compute the correlation function of the XXZ model
\cite{CBMS,DFJMN}.

Similarly there is a model corresponding to
the elliptic (genus-$1$) solution
of the Yang-Baxter equation called the XYZ model.
This is defined in terms of the elliptic algebra
$\mathcal{A}_{q,p}(\widehat{sl}_2)$ \cite{FIJKMY1,FIJKMY2}.
As an analogy of quantum affine algebra
the VO attached to the elliptic algebra is also introduced.
By using the elliptic VO
the difference equations and the integral expressions
for correlation functions are found \cite{JMN,LP}.

The vertex operator approach is valid for the half-infinite chains
with a boundary;
the VO indicates the half-infinite spin chain and
the transfer matrix is then composed by the VOs \cite{JKKKM1,JKKKM2}.
In this formulation
the correlation function is shown to obey a $q$-difference equation
called the boundary q-KZ equation,
similar to the quantized Knizhnik-Zamolodchikov (q-KZ) equation \cite{FR,S}.
The magnetization is calculated by solving it.

It has been shown that
the eigenstates in the finite XXZ chain with boundaries
can be defined by using the VOs \cite{Fujii1,Fujii2}
although the VO indicates the half-infinite chain.
There, the transfer matrix for the bounded chain
is derived from the boost operator of the boundary q-KZ equation.
Furthermore, by noting the geometric symmetry of the model,
the spontaneous magnetization in the thermodynamic limit has been calculated,
and a relation for magnetizations in the bounded chain
and in the half-infinite chain has been obtained.
This relation enables us to compute the magnetization in the bounded chain
from that in the half-infinite chain.
We here remark that the VO has been already extended to the elliptic case.
One may construct the eigenstates in the XYZ model
by the elliptic VOs in the same way as the XXZ model
if there exists a similar relation for magnetizations
in the XYZ model.

In the paper we generalize the relation in the XXZ model
and show the ``finite size formula'' in the bounded XYZ model
(see (\ref{dual})).
We study the following XYZ Hamiltonian with boundary magnetic fields,
\begin{equation}
  \label{h}
  H_{\mbox{\scriptsize XYZ}} =
  -\frac{1}{2}\sum_{i=1}^{N-1}
  ((1+\Gamma)\sigma_i^x\sigma_{i+1}^x+(1-\Gamma)\sigma_i^y\sigma_{i+1}^y
  +\Delta\sigma_i^z\sigma_{i+1}^z)+f_1\sigma_1^z+f_N\sigma_N^z.
\end{equation}
The anisotropic parameters $\Gamma$, $\Delta$ are related with
$p$, $q$ and
the boundary magnetic fields $f_1$, $f_N$ are
determined by an extra parameter $r$ (see (\ref{gd}) and (\ref{f1})).
In Section~\ref{vo} we introduce the elliptic VO
and express the eigenstates of the model in terms of the VOs.
We also introduce the idea of the turning symmetry,
which is a geometric symmetry
in the bounded XYZ spin chain without magnetic fields.
In Section~\ref{dfi} we consider two infinite chain limits;
the half-infinite chain limit and the thermodynamic limit.
In the former one of the boundaries vanishes
and in the latter the both boundaries remain.
We formulate the magnetizations in the both cases
and derive the finite size formula (\ref{dual}).
In Section~\ref{sp}, using this formula,
we calculate the spontaneous magnetization
in the bounded XYZ model.

\setcounter{equation}{0}
\section{Vertex Operators and Turing Symmetry} 
\label{vo}
\subsection{Vertex Operator Expressions of Vacuum States}
We define the VO in the level one modules of
the elliptic algebra $\mathcal{A}_{q,p}(\widehat{sl}_2)$ 
to be the intertwiner with a map
\begin{equation}
  \Phi^{(1-i,i)}(\zeta) :
  \mathcal{F}^{(i)}\rightarrow \mathcal{F}^{(1-i)}\otimes V,
\end{equation}
where $\mathcal{F}^{(i)}$ $(i=0,1)$ are the half-infinite paths
\begin{equation}
  \label{path}
  \mathcal{F}^{(i)} =
  v_{p(1)}\otimes\cdots\otimes v_{p(N)}\otimes\cdots,
  \qquad
  p(k) = (-)^{k+i},
  \qquad (k\gg 1)
\end{equation}
and $V=\mathbb{C}v_+\oplus\mathbb{C}v_-$ is the evaluation module.
Paths $\otimes_{k=1}^\infty v_{(-)^{k+i}}$ ($i=0,1$)
are called the ground states.
The VO obeys the following commutation relation,
\begin{equation}
  \label{com}
  R_{1,2}(\zeta_1/\zeta_2)
  \Phi_1^{(i,1-i)}(\zeta_1)\Phi_2^{(1-i,i)}(\zeta_2) =
  \Phi_2^{(i,1-i)}(\zeta_2)\Phi_1^{(1-i,i)}(\zeta_1).
\end{equation}
The R-matrix $R_{i,j}(\zeta)$ is
the intertwiner in $\mathcal{A}_{q,p}(\widehat{sl}_2)$
with a map $\mbox{End}_\mathbb{C}(V_i\otimes V_j)$
and satisfies the Yang-Baxter equation (see (\ref{ybe})).

Let us construct the eigenstates of
the bounded XYZ model.
We need not only the VOs but the boundary states
$|W\rangle^{(i)}\in\mathcal{F}^{(i)}$
and ${}^{(i)}\langle V|\in\mathcal{F}^{*(i)}$
that satisfy the following defining relations,
\begin{eqnarray}
  \label{w}
  K(\zeta;r)\Phi^{(1-i,i)}(\zeta)|W\rangle^{(i)} &=&
  \Lambda_W^{(i)}(\zeta;r)\times\Phi^{(1-i,i)}(\zeta^{-1})|W\rangle^{(i)},
  \\
  \label{v}
  {}^{(i)}\langle V|\Phi^{(i,1-i)}(\zeta^{-1})K(\zeta;\bar{r}) &=&
  \Lambda_V^{(i)}(\zeta;\bar{r})\times
  {}^{(i)}\langle V|\Phi^{(i,1-i)}(\zeta).
\end{eqnarray}
Here $K(\zeta;r)$ is the boundary K-matrix for the XYZ model
with a map $\mbox{End}_\mathbb{C}(V)$ \cite{Sk}.
Factors $\Lambda_W^{(i)}(\zeta;r)$ and $\Lambda_V^{(i)}(\zeta;\bar{r})$
are determined later (see (\ref{l})).
With the boundary states $|W\rangle^{(i)}$ and ${}^{(i)}\langle V|$
we express the eigenstates in the XYZ spin chain by
\begin{equation}
  \label{z}
  |\zeta_1,\ldots,\zeta_N\rangle^{(i)} =
  {}^{(i)}\langle V|\Phi^{(i,1-i)}(\zeta_1)\otimes\cdots\otimes
  \Phi^{(1-i,i)}(\zeta_N)|W\rangle^{(i)},
\end{equation}
where we have assumed that the number of sites is even for simplicity.
It is obvious that the state (\ref{z}) is
indeed the eigenstate for the model:\
from the commutation relation (\ref{com})
and the defining relations (\ref{w}), (\ref{v}),
it can be deduced that
\begin{equation}
  \label{tz}
  T(\zeta_1,\ldots,\zeta_N)|\zeta_1,\ldots,\zeta_N\rangle^{(i)} =
  \Lambda_W^{(i)}(\zeta;r)\Lambda_V^{(i)}(\zeta;\bar{r})
  |\zeta_1,\ldots,\zeta_N\rangle^{(i)},
\end{equation}
with the boost operator
\begin{eqnarray}
  \label{t}
  T(\zeta_1,\ldots,\zeta_N)
  &=&
  R_{N,N-1}(\zeta_N/\zeta_{N-1})\cdots R_{N,1}(\zeta_N/\zeta_1)
  K_N(\zeta_N;\bar{r})
  \nonumber \\ && \times
  R_{1,N}(\zeta_1 \zeta_N)\cdots R_{N-1,N}(\zeta_{N-1}\zeta_N)
  K_N(\zeta_N;r).
\end{eqnarray}
Since $T(1,\ldots,1)=1$ and
$\Lambda_W^{(i)}(1;r)=\Lambda_V^{(i)}(1;\bar{r})=1$ (see (\ref{l})),
differentiating (\ref{tz}) at $\zeta_1,\ldots,\zeta_N=1$ yields
\begin{equation}
  \left.\frac{\partial}{\partial\zeta_N}T(\zeta_1,\ldots,\zeta_N)
  \right|_{\zeta_1,\ldots,\zeta_N=1}
  \hspace{-1em}\times|1,\ldots,1\rangle^{(i)} \propto
  |1,\ldots,1\rangle^{(i)}.
\end{equation}
We then show that the boost operator (\ref{t}) is
the transfer matrix for the XYZ model.
Let $h_{i,i+1}$ be the local element of the Hamiltonian
on the site $(i,i+1)$.
The following relations
\begin{equation}
  \left.\frac{\partial}{\partial\zeta}R_{i,i+1}(\zeta)\right|_{\zeta=1}
  \propto h_{i,i+1}P_{i,i+1}+\mbox{const.},
  \qquad
  \left.\frac{\partial}{\partial\zeta}K_i(\zeta;r)\right|_{\zeta=1}
  \propto f_i\sigma_i^z+\mbox{const.},
\end{equation}
implies that
the derivative of the boost operator generates the XYZ Hamiltonian,
\begin{equation}
  \left.\frac{\partial}{\partial\zeta_N}
    T(\zeta_1,\ldots,\zeta_N)\right|_{\zeta_1,\ldots,\zeta_N=1} \propto
  H_{\mbox{\scriptsize XYZ}}+\mbox{const.}.
\end{equation}
Here $P_{i,j}$ is the permutator:
$V_i\otimes V_j\rightarrow V_j\otimes V_i$.
Thus, the boost operator (\ref{t}) is regarded as
the transfer matrix in the XYZ model
and the state (\ref{z}) is the eigenstate of the XYZ Hamiltonian.

The dual state is given as above
by using the dual VO $\Phi^{*(1-i,i)}(\zeta)=\Phi^{(1-i,i)}(-q^{-1}\zeta)$,
\begin{equation}
  \label{dz}
  {}^{(i)}\langle\zeta_1,\ldots,\zeta_N| =
  {}^{(i)}\langle W^*|
  \Phi^{*(i,1-i)}(\zeta_N)\otimes\cdots\otimes\Phi^{*(1-i,i)}(\zeta_1)
  |V^*\rangle^{(i)},
\end{equation}
with the boundary states ${}^{(i)}\langle W^*|\in\mathcal{F}^{*(i)}$
and $|V^*\rangle^{(i)}\in\mathcal{F}^{(i)}$
being defined by the relations,
\begin{eqnarray}
  \label{dw}
  {}^{(i)}\langle W^*|\Phi^{*(i,1-i)}(\zeta^{-1})K(\zeta;r) &=&
  \Lambda_W^{(i)}(\zeta;r)\times
  {}^{(i)}\langle W^*|\Phi^{*(i,1-i)}(\zeta),
  \\
  \label{dv}
  K(\zeta;\bar{r})\Phi^{*(1-i,i)}(\zeta)|V^*\rangle^{(i)} &=&
  \Lambda_V^{(i)}(\zeta;\bar{r})\times
  \Phi^{*(1-i,i)}(\zeta^{-1})|V^*\rangle^{(i)}.
\end{eqnarray}
By definition the dual state obeys
\begin{equation}
  {}^{(i)}\langle\zeta_1,\ldots,\zeta_N|T(\zeta_1,\ldots,\zeta_N) =
  {}^{(i)}\langle\zeta_1,\ldots,\zeta_N|
  \Lambda_W^{(i)}(\zeta;r)\Lambda_V^{(i)}(\zeta;\bar{r}).
\end{equation}
For the XXZ model the states
constructed by the VOs like (\ref{z}) and (\ref{dz})
are known to be the vacuum states \cite{Fujii2}.
We assume that the states (\ref{z}) and (\ref{dz})
correspond to the vacuum states in the XYZ model.

We point out a similarity between the transfer matrix (\ref{t})
and the boost operator of a boundary q-KZ equation \cite{Fujii2}.
The boundary q-KZ equation is a difference equation like (\ref{tz})
that the correlation function in the half-infinite XYZ spin chain satisfies.
(In the boundary q-KZ equation ``$|\zeta_1,\ldots,\zeta_N\rangle$''
play a role of the correlation function.)
Its solution is given in terms of the VOs \cite{JKKKM2}.
In the bounded XYZ spin chain,
making use of such a similarity,
we have successfully constructed the vacuum states
by the vertex operator approach.

\subsection{Turning Symmetry}
The spontaneous magnetization is generated
in the states with no boundary magnetic fields.
Notice that such states are invariant when they are turned over.
We call this geometric symmetry the \textit{turning symmetry}
and consider only the vacuum states satisfying this symmetry.
Since a parameter $r$ is inverted
when the corresponding boundary magnetic field is reversed
(see (\ref{f1})),
the following relation holds:
\begin{eqnarray}
  \label{turn}
  \lefteqn{{}^{(i)}\langle V(\bar{r})|
  \Phi^{(i,1-i)}(\zeta_1)\otimes\cdots\otimes\Phi^{(1-i,i)}(\zeta_N)
  |W(r)\rangle^{(i)}}
  \nonumber \\ &\hookrightarrow&
  {}^{(i)}\langle V(r^{-1})|
  \Phi^{(i,1-i)}(\zeta_N)\otimes\cdots\otimes\Phi^{(1-i,i)}(\zeta_1)
  |W(\bar{r}^{-1})\rangle^{(i)}
  \nonumber \\ &\simeq&
  {}^{(i)}\langle V(r^{-1})|
  \Phi^{(i,1-i)}(\zeta_1)\otimes\cdots\otimes\Phi^{(1-i,i)}(\zeta_N)
  |W(\bar{r}^{-1})\rangle^{(i)}.
\end{eqnarray}
We have denoted the parameter $r$ dependence explicitly.
This relation implies that
\begin{equation}
  \label{r}
  \bar{r} = r^{-1}.
\end{equation}
The turning symmetry plays a central role
in the analysis of the bounded XYZ model.

\setcounter{equation}{0}
\section{Finite Size Formula in the Infinite Chain Limits}
\label{dfi}
\subsection{Half-Infinite Chain Limit}
It is important in physics to consider the infinite chain limit.
First we consider the half-infinite chain limit,
which means that one of boundaries vanishes.
The vacuum states are then expressed by
\begin{eqnarray}
  \label{iz}
  |\ldots,\zeta_N,\ldots,\zeta_1\rangle^{(i)} &=&
  \cdots\otimes
  \Phi(\zeta_N)\otimes\cdots\otimes\Phi(\zeta_1)
  |W\rangle^{(i)},
  \\
  \label{idz}
  {}^{(i)}\langle\zeta_1,\ldots,\zeta_N,\ldots| &=&
  {}^{(i)}\langle W^*|
  \Phi^*(\zeta_1)\otimes\cdots\otimes\Phi^*(\zeta_N)
  \otimes\cdots.
\end{eqnarray}
Hereafter the label $(1-i,i)$ on the VO is omitted
unless this omission induces confusion.

By using the states (\ref{iz}) and (\ref{idz})
let us calculate the vacuum expectation value of
an operator $\mathcal{O}$ on the site $m$.
We define its action by
$\mathcal{O}.v_\epsilon =
\sum_{\epsilon'=\pm}\mathcal{O}_\epsilon^{\epsilon'}v_{\epsilon'}$.
Decomposing the VOs into the vector forms
\begin{equation}
  \Phi(\zeta) = \sum_{\epsilon=\pm}\Phi_\epsilon(\zeta).v_\epsilon,
  \qquad
  \Phi^*(\zeta)=\sum_{\epsilon=\pm}\Phi_\epsilon^*(\zeta).v_\epsilon^*,
\end{equation}
and taking care of the unitarity relation \cite{FIJKMY1,FIJKMY2}
\begin{equation}
  \label{uni}
  \Phi^*(\zeta).\Phi(\zeta) = \mbox{id},   
\end{equation}
we get the following vacuum expectation value of $\mathcal{O}$,
\begin{eqnarray}
  \label{o}
  \langle\mathcal{O}\rangle_\infty^{(i)} &=&
  \sum_{\epsilon_1,\ldots,\epsilon_m,\epsilon'_m=\pm}
  {}^{(i)}\langle W^*|
  \Phi_{\epsilon_1}^*(\zeta_1)\cdots \Phi_{\epsilon_m}^*(\zeta_m)
  \mathcal{O}_{\epsilon_m}^{\epsilon'_m}
  \Phi_{\epsilon'_m}(\zeta_m)\cdots \Phi_{\epsilon_1}(\zeta_1)
  |W\rangle^{(i)}.
\end{eqnarray}
The attached subscript $\infty$ indicates the half-infinite chain limit.
Apart from a normalization
this expectation value is equal to that already obtained in \cite{JKKKM2}.
The magnetization at a boundary is then written by
\begin{equation}
  \label{mi}
  \langle\sigma^z\rangle_\infty^{(i)} =
  \frac{P_+^{(i)}(-q^{-1},1)-P_-^{(i)}(-q^{-1},1)}
  {P_+^{(i)}(-q^{-1},1)+P_-^{(i)}(-q^{-1},1)},
\end{equation}
with
\begin{equation}
  \label{p}
  P_\epsilon^{(i)}(\zeta_1,\zeta_2) =
  {}^{(i)}\langle W^*|
  \Phi_{-\epsilon}(\zeta_1)\Phi_\epsilon(\zeta_2)
  |W\rangle^{(i)}.
\end{equation}
In return, in order to deduce the expectation value (\ref{o}),
we have defined the dual state (\ref{dz}) by using the dual VOs.
This $P_\epsilon^{(i)}(\zeta_1,\zeta_2)$ obeys
the boundary q-KZ equations (\ref{bqkz1}) and (\ref{bqkz2}).
By solving these equations
$P_\epsilon^{(i)}(\zeta_1,\zeta_2)$ is determined.

\subsection{Baxter's Formula}
Next we consider the thermodynamic limit
such that the both boundaries survive.
Due to a boundary state ${}^{(i)}\langle V|$
one can not utilize the unitarity relation (\ref{uni}).
To deal with the infinite product of the VOs directly,
we interpret the VO as the half-infinite product of the R-matrices,
\begin{equation}
  \label{f}
  \Phi^{(1-i,i)}(\zeta) =
  \widehat{R}_{1,2}(\zeta)\cdots\widehat{R}_{N,N+1}(\zeta)\cdots,
\end{equation}
where we have set $\widehat{R}_{j,k}(\zeta)=R_{j,k}(\zeta)P_{j,k}$.
The R-matrices for sufficiently large $N$
describe the ground state $\otimes_{k=1}^\infty v_{(-)^{k+i}}$.

From the interpretation (\ref{f})
some formulae for the asymptotic form the VO are derived \cite{CBMS}.
We start from the Yang-Baxter equation
\begin{equation}
  \label{ybe}
  \widehat{R}_{j,j+1}(\zeta)\widehat{R}_{j+1,j+2}(\xi\zeta)
  \widehat{R}_{j,j+1}(\xi) =
  \widehat{R}_{j+1,j+2}(\xi)\widehat{R}_{j,j+1}(\xi\zeta)
  \widehat{R}_{j+1,j+2}(\zeta).
\end{equation}
Differentiating it at $\xi=1$ yields
\begin{eqnarray}
  \lefteqn{\widehat{R}_{j,j+1}(\zeta)\widehat{R}_{j+1,j+2}(\zeta)h_{j,j+1}
    -h_{j+1,j+2}\widehat{R}_{j,j+1}(\zeta)\widehat{R}_{j+1,j+2}(\zeta)}
  \nonumber \\ &=&
  \zeta\frac{d\widehat{R}_{j,j+1}(\zeta)}{d\zeta}
  \widehat{R}_{j+1,j+2}(\zeta)
  -\widehat{R}_{j,j+1}(\zeta)
  \zeta\frac{d\widehat{R}_{j+1,j+2}(\zeta)}{d\zeta}.
\end{eqnarray}
Multiply $j\widehat{R}_{1,2}(\zeta)\cdots\widehat{R}_{j-1,j}(\zeta)$
from the left and
$\widehat{R}_{j+2,j+3}(\zeta)\cdots\widehat{R}_{N-1,N}(\zeta)\cdots$
from the right.
Summing it over $j=1,\ldots,N,\ldots$ we get
\begin{eqnarray}
  \lefteqn{(\widehat{R}_{1,2}(\zeta)\cdots\widehat{R}_{N,N+1}(\zeta)\cdots)
  (h_{1,2}+2h_{2,3}+\cdots+Nh_{N,N+1}+\cdots)}
  \nonumber \\ && -
  (h_{1,2}+2h_{2,3}+\cdots+Nh_{N,N+1}+\cdots)
  (\widehat{R}_{1,2}(\zeta)\cdots\widehat{R}_{N,N+1}(\zeta)\cdots)
  \nonumber \\ &=&
  \zeta\frac{d}{d\zeta}
  (\widehat{R}_{1,2}(\zeta)\cdots\widehat{R}_{N,N+1}(\zeta)\cdots).
\end{eqnarray}
Let $D=\sum_{j=1}^\infty jh_{j,j+1}$.
By integrating this equation over $\zeta$ the following formula is deduced,
\begin{equation}
  \label{hom}
  \xi^{-D}\Phi(\zeta)\xi^D = \Phi(\zeta/\xi).
\end{equation}
Further Baxter discovered that the corner transfer matrix
\begin{equation}
  A(\zeta) =
  (\widehat{R}_{1,2}(\zeta)\cdots\widehat{R}_{N,N+1}(\zeta)\cdots)
  \otimes\cdots\otimes
  (\widehat{R}_{N,N+1}(\zeta)\cdots)\otimes\cdots
\end{equation}
is simply expressed as
\begin{equation}
  \label{bf}
  A(\zeta)=\zeta^{-D},
\end{equation}
owing to ignoring contributions of the ground state \cite{B}.
We call this the Baxter's formula.

\begin{figure}[t]
  \begin{center}
    \psfrag{w}{$|W\rangle$}
    \psfrag{=}{$=$}
    \includegraphics[width=13cm]{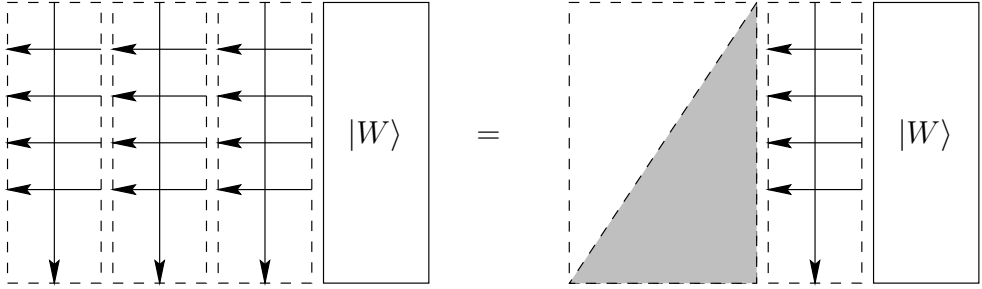}
    \caption{\small The Baxter formula simplifies the expression
      of the vacuum state.
      The direction of the horizon arrows corresponds to the order
      of the spectral parameters.}
    \label{bax}
  \end{center}
\end{figure}

The formula (\ref{hom}) and 
the Baxter's formula (\ref{bf}) lead us to
the asymptotic form of the infinite product of the VOs as follows.
The R-matrices belonging to the VOs
at sufficiently far sites are in the ground state,
in the same manner as very far R-matrices of the VO
are in the ground state.
Therefore, the upper triangle sector in the huge products of the R-matrices,
or the infinite product of the VOs, is certainly in the ground state
(Figure~\ref{bax}).
Since there is no contribution of the ground state in the Baxter's formula,
we have
\begin{equation}
  \cdots\Phi(\zeta_B)\otimes\Phi(\zeta_B)\Phi_\epsilon(\zeta)
  |W\rangle^{(i)} =
  \zeta_B^{-D}\Phi_\epsilon(\zeta)|W\rangle^{(i)},
\end{equation}
where $\zeta_B$ stands for the spectral parameter in the balk part.
The VO near the boundary state remains
because the VO can not be decomposed into the R-matrices
for the sake of the defining relation (\ref{w}).

\begin{figure}[t]
  \begin{center}
    \psfrag{v}{$\langle V|$}
    \psfrag{w}{$|W\rangle$}
    \psfrag{=}{$=$}
    \includegraphics[width=10cm]{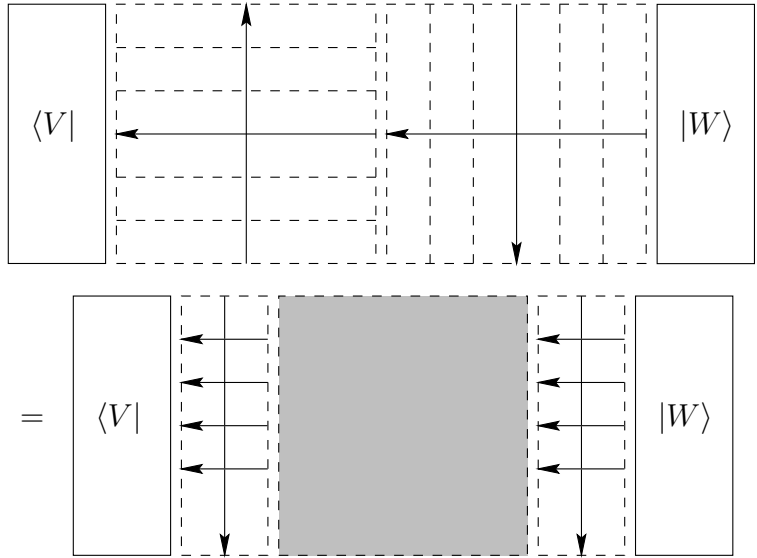}
    \caption{\small The turning symmetry is realized by
      the product of the R-matrices and the $90^\circ$-rotated matrices.
      Note that the direction of the horizon arrow corresponds to
      the order of the spectral parameters
      and it is reversible (see the last line of (\ref{turn})).}
    \label{rot}
  \end{center}
\end{figure}

Let us consider the VOs near another boundary.
Note that the states satisfy the turning symmetry.
They are realized as products of
R-matrices and $90^\circ$-rotated matrices,
which are given by the crossing symmetry
$\widehat{R}(\zeta)^{rot}=\widehat{R}(-q^{-1}\zeta^{-1})$.
(The symbol $rot$ means the rotation.)
Hence, noting the relation (\ref{hom}),
we can simplify the vacuum state in the thermodynamic limit as follows
(Figure~\ref{rot}),
\begin{eqnarray}
  \label{sz}
  |\zeta,\zeta_B\rangle_\epsilon^{(i)} &=&
  {}^{(i)}\langle V|
  (-q^{-1}\zeta_B^{-1})^{-D}\Phi_{-\epsilon}(-q^{-1}\zeta)
  \times\zeta_B^{-D}\Phi_\epsilon(\zeta)
  |W\rangle^{(i)}
  \nonumber \\ &=&
  {}^{(i)}\langle V|
  \Phi_{-\epsilon}(\zeta\zeta_B)(-q)^D \Phi_\epsilon(\zeta)
  |W\rangle^{(i)}.
\end{eqnarray}
Here we use the rotated VO related with
$\Phi_\epsilon(\zeta)^{rot}=\Phi_{-\epsilon}(-q^{-1}\zeta)$.
We have obtained a compact expression (\ref{sz}) for the vacuum state.

To verify this simplification
we consider the infinite chain without boundaries.
The vacuum state of such chain clearly satisfies the turning symmetry
and is expressed by the infinite product of the VOs.
In the same way as above we have
\begin{equation}
  |\zeta_B\rangle = (-q^{-1}\zeta_B^{-1})^{-D}\times\zeta_B^{-D} = (-q)^D.
\end{equation}
This is equal to the vacuum state obtained
by the original methods based on
the corner transfer matrix \cite{CBMS,JMN,B}.

Similarly the dual vacuum state is reduced to
\begin{equation}
  \label{sdz}
  {}_{\;\;\epsilon}^{(i)}\langle\zeta,\zeta_B| =
  {}^{(i)}\langle W^*|
  \Phi_\epsilon^*(\zeta\zeta_B)(-q)^D\Phi_{-\epsilon}^*(\zeta)
  |V^* \rangle^{(i)}.
\end{equation}
With these expressions, in the thermodynamic limit,
the magnetization of the bounded chain is written by
\begin{equation}
  \label{m}
  \langle\sigma^z\rangle^{(i)} =
  \frac{{}_{\;+}^{(i)}\langle 1,1|1,1\rangle_+^{(i)}
    -{}_{\;-}^{(i)}\langle 1,1|1,1\rangle_-^{(i)}}
  {{}_{\;+}^{(i)}\langle 1,1|1,1\rangle_+^{(i)}
    +{}_{\;-}^{(i)}\langle 1,1|1,1\rangle_-^{(i)}}.
\end{equation}
This is the magnetization at a boundary.
Using the fact that $\Phi(1)$ corresponds to
the translation operator (recall (\ref{f}) and give care to $R(1)=P$),
one can obtain the magnetization at any site.

\subsection{Finite Size Formula}

One may find that the simplified expressions of the states
(\ref{sz}) and (\ref{sdz})
are similar to $P_\epsilon^{(i)}(\zeta_1,\zeta_2)$
appeared in the half infinite chain limit (\ref{p}).
We show that they are indeed the same.

In the half-infinite XYZ chain \cite{JKKKM2},
through the defining relations (\ref{w}) and (\ref{dw})
for $|W\rangle^{(i)}$ and ${}^{(i)}\langle W^*|$,
the factor $\Lambda_W^{(i)}(\zeta;r)$ is determined by
\begin{equation}
  \Lambda_W^{(0)}(\zeta;r) = 1,
  \qquad
  \Lambda_W^{(1)}(\zeta;r) =
  \Lambda(\zeta;r) =
  \frac{\Theta_{q^4}(r\zeta^2)\Theta_{q^4}(q^2 r\zeta^{-2})}
  {\Theta_{q^4}(r\zeta^{-2})\Theta_{q^4}(q^2 r\zeta^2)}\zeta^{-2},
\end{equation}
where
\begin{equation}
  \Theta_t(z) = (z;t)_\infty(tz^{-1};t)_\infty(t;t)_\infty,
\end{equation}
with
\begin{equation}
  (z;t_1,\ldots,t_k)_\infty =
  \prod_{n_1,\ldots,n_k=0}^\infty(1-zt_1^{n_1}\cdots t_k^{n_k}).
\end{equation}
With this $\Lambda(\zeta;r)$,
from the explicit expression of the boundary K-matrix (\ref{k}),
the following relation is derived,
\begin{equation}
  K_\epsilon^\epsilon(\zeta;r) =
  \Lambda(\zeta;r)K_{-\epsilon}^{-\epsilon}(\zeta;r^{-1}).
\end{equation}
Applying this relation to the defining relation of the boundary states
(\ref{w}), (\ref{v}), (\ref{dw}) and (\ref{dv}),
and using the turning symmetric relation (\ref{r}),
one can get the following identities,
\begin{equation}
  \label{l}
  \Lambda_W^{(0)}(\zeta;r) = \Lambda_V^{(1)}(\zeta;r) = 1,
  \qquad
  \Lambda_W^{(1)}(\zeta;r) = \Lambda_V^{(0)}(\zeta;r) = \Lambda(\zeta;r),
\end{equation}
and
\begin{equation}
  \label{nov}
  {}^{(i)}\langle V| = {}^{(i)}\langle W^*|(-q)^{-D},
  \qquad
  |V^*\rangle^{(i)} = (-q)^{-D}|W\rangle^{(i)}.
\end{equation}
By the identities (\ref{nov}),
the states (\ref{sz}) and (\ref{sdz}) are related to
$P_\epsilon^{(i)}(\zeta_1,\zeta_2)$ in (\ref{p}) as
\begin{equation}
  \label{zp}
  |\zeta_2,\zeta_1/\zeta_2\rangle_\epsilon^{(i)} =
  {}_{\;\;\epsilon}^{(i)}\langle\zeta_2,\zeta_1/\zeta_2| =
  P_\epsilon^{(i)}(-q^{-1}\zeta_1,\zeta_2).
\end{equation}
Recall the expressions of the magnetizations (\ref{mi}) and (\ref{m}).
From the identification (\ref{zp})
we conclude the following formula in the bounded XYZ model:
\begin{equation}
  \label{dual}
  \langle\sigma^z\rangle^{(i)} =
  \frac{2\langle\sigma^z\rangle_\infty^{(i)}}
  {1+(\langle\sigma^z\rangle_\infty^{(i)})^2}.
\end{equation}
We call (\ref{dual}) the \textit{finite size formula}.
The finite size formula,
which relates the magnetization in the bounded model
with that in the half-infinite model,
generally holds
in the spin chains satisfying the turning symmetry,
for example, the XXZ model with boundaries \cite{Fujii2}.

Let us consider on the q-KZ equation.
Originally, the half-infinite XYZ spin chain was formulated
by the VOs based on the corner transfer matrix method,
and its correlation function was shown to
obey the boundary q-KZ equation as (\ref{tz}) \cite{JKKKM2}.
By solving the boundary q-KZ equation
the two-point function is computed.
On the other hand, making use of a similarity between
the transfer matrix (\ref{t}) and
the boost operator of the boundary q-KZ equation,
we have constructed the vacuum state by the vertex operator approach.
The roles of the q-KZ equation are clearly different.
However, discovering the turning symmetry,
we have obtained similar two-point function in the bounded model.
There, the boundary q-KZ equation is again useful
for calculating the two-point function in the bounded model.
It can be said that,
by virtue of the geometric symmetry of the model,
the q-KZ equation retrieves its original roles
and yields the finite size formula (\ref{dual}).
This formula gives us the magnetization
in the bounded chain by that in the half-infinite chain.

\setcounter{equation}{0}
\section{Spontaneous Magnetizations in the XYZ Model}
\label{sp}
\subsection{The XYZ Model}
Let us calculate the spontaneous magnetization in the XYZ model.
We introduce the following parameters,
\begin{equation}
  \label{pqzr}
  p = e^{-\frac{\pi K'}{K}},
  \qquad
  -q = e^{-\frac{\pi\lambda}{2K}},
  \qquad
  \zeta = e^{\frac{\pi u}{2K}},
  \qquad
  r = e^{\frac{\pi\alpha}{K}},
\end{equation}
where $K$ and $K'$ are the complete elliptic integrals
of the first kind.
The region of parameters is restricted to
$0<p^{1/2}<-q<|r|^{1/2}<|\zeta|^{-1}<1$
for the boundary condition of paths (\ref{path}).
We use both ``multiplicative'' parameters $p$, $q$, $\zeta$, $r$
and ``additive'' parameters $\lambda$, $u$, $\alpha$
for the elliptic nome $p$ on an equal footing.
We define the actions of the R-matrix and K-matrix by
\begin{eqnarray}
  R(\zeta).(v_{\epsilon_1}\otimes v_{\epsilon_2}) &=&
  \sum_{\epsilon'_1,\epsilon'_2=\pm}
  R_{\epsilon_1\epsilon_2}^{\epsilon'_1\epsilon'_2}(\zeta)
  (v_{\epsilon'_1}\otimes v_{\epsilon'_2}),
  \\
  K(\zeta;r).v_\epsilon &=&
  \sum_{\epsilon'=\pm}K_\epsilon^{\epsilon'}(\zeta;r)v_{\epsilon'}.  
\end{eqnarray}
Then the non-zero matrix elements of R-matrix are
\begin{eqnarray}
  R_{++}^{++}(\zeta) = R_{--}^{--}(\zeta) &=&
  \mbox{snh}(\lambda-u)\gamma(\zeta),
  \\
  R_{+-}^{+-}(\zeta) = R_{-+}^{-+}(\zeta) &=&
  \mbox{snh}(u)\gamma(\zeta),
  \\
  R_{-+}^{+-}(\zeta) = R_{+-}^{-+}(\zeta) &=&
  \mbox{snh}(\lambda)\gamma(\zeta),
  \\
  R_{--}^{++}(\zeta) = R_{++}^{--}(\zeta) &=&
  k\mbox{snh}(\lambda-u)\mbox{snh}(u)\mbox{snh}(\lambda)\gamma(\zeta),
\end{eqnarray}
where $k=4p^{1/2}(-p^2;p^2)_\infty^4/(-p;p^2)_\infty^4$
is the modulus of elliptic functions
\begin{equation}
  \mbox{snh}(u) = -i\mbox{sn}(iu),
  \qquad
  \mbox{cnh}(u) = \mbox{cn}(iu),
  \qquad
  \mbox{dnh}(u) = \mbox{dn}(iu),
\end{equation}
and
\begin{equation}
  \gamma(\zeta) = 
  \frac{(p^2;p^2)_\infty}{(p;p)_\infty^2}
  \frac{\Theta_{p^2}(q^2)\Theta_{p^2}(p\zeta^2)}{\Theta_{p^2}(q^2\zeta^2)}
  \frac{g(\zeta^2)}{\mbox{snh}(\lambda)},
\end{equation}
\begin{equation}
  g(z) =
  \frac{(q^2 z;p,q^4)_\infty(q^4 z^{-1};p,q^4)_\infty
    (pz^{-1};p,q^4)_\infty(pq^2 z;p,q^4)_\infty}
  {(q^2 z^{-1};p,q^4)_\infty(q^4 z;p,q^4)_\infty
    (pz;p,q^4)_\infty(pq^2 z^{-1};p,q^4)_\infty}.
\end{equation}
In the limit $p\rightarrow 0$ the model reduces to the XXZ model.

The boundary K-matrix is given by \cite{IK}
\begin{equation}
  \label{k}
  K_+^+(\zeta;r) =
  \frac{\mbox{snh}(\alpha+u)}{\mbox{snh}(\alpha-u)}
  \frac{\kappa(\zeta^2;r)}{\kappa(\zeta^{-2};r)},
  \qquad
  K_-^-(\zeta;r) =
  \frac{\kappa(\zeta^2;r)}{\kappa(\zeta^{-2};r)},
\end{equation}
with
\begin{eqnarray}
  \kappa(z;r) &=&
  \frac{(prz;p^2,q^4)_\infty(p^2 r^{-1}z;p^2,q^4)_\infty
    (rq^4 z;p^2,q^4)_\infty(pr^{-1}q^4 z;p^2,q^4)_\infty}
  {(prq^2 z;p^2,q^4)_\infty(p^2r^{-1}q^2 z;p^2,q^4)_\infty
    (rq^2 z;p^2,q^4)_\infty(pr^{-1}q^2 z;p^2,q^4)_\infty}
  \nonumber \\ && \times
  \frac{(q^6 z^2;p^2,q^8)_\infty(pq^2 z^2;p^2,q^8)_\infty
    (pq^6 z^2;p^2,q^8)_\infty(p^2 q^2 z^2;p^2,q^8)_\infty}
  {(q^8 z^2;p^2,q^8)_\infty(pq^4 z^2;p^2,q^8)_\infty^2
    (p^2 z^2;p^2,q^8)_\infty}.
\end{eqnarray}
From these explicit expressions of R-matrix and K-matrix,
the anisotropic parameters and the boundary magnetic fields
of the XYZ Hamiltonian (\ref{h}) are determined by
\begin{equation}
  \label{gd}
  \Gamma = k\mbox{snh}^2(\lambda),
  \qquad
  \Delta = -\mbox{cnh}(\lambda)\mbox{dnh}(\lambda),
\end{equation}
\begin{equation}
  \label{f1}
  f_1 = -f_N =
  \frac{\mbox{snh}(\lambda)\mbox{cnh}(\alpha)\mbox{dnh}(\alpha)}
  {2\mbox{snh}(\alpha)}.
\end{equation}
Here we have used the turning symmetric relation (\ref{r}).
Recall that $\lambda$ and $\alpha$ are related to $p$, $q$ and $r$
by (\ref{pqzr}).
The boundary magnetic fields vanish when $r=-1$.

\subsection{Spontaneous Magnetizations}
To calculate the spontaneous magnetization,
we put $r=-1$ and consider a function coupled with vectors,
\begin{equation}
  P(\zeta_1,\zeta_2) =
  \sum_{\epsilon=\pm}P_\epsilon^{(i)}(\zeta_1,\zeta_2).
  (v_{-\epsilon}\otimes v_\epsilon).
\end{equation}
Here and hereafter the label $i$ is omitted
because there is no difference for $i=0$ and $1$;\
$\Lambda^{(0)}(\zeta;-1)=\Lambda^{(1)}(\zeta;-1)=1$.
This function obeys the boundary q-KZ equations
\begin{eqnarray}
  \label{bqkz1}
  P(q^{-2}\zeta_1,\zeta_2) &=&
  K_1(-q^{-1}\zeta_1;-1)R_{21}(\zeta_1\zeta_2)K_1(\zeta_1;-1)
  R_{12}(\zeta_1/\zeta_2)P(\zeta_1,\zeta_2),
  \\
  \label{bqkz2}
  P(\zeta_1,q^{-2}\zeta_2) &=&
  R_{21}(q^{-2}\zeta_2/\zeta_1)K_2(-q^{-1}\zeta_2;-1)
  R_{12}(\zeta_1\zeta_2)K_2(\zeta_2;-1)
  P(\zeta_1,\zeta_2).
\end{eqnarray}
By use of new functions,
\begin{equation}
  F_\pm(\zeta_1,\zeta_2) = P_-(\zeta_1,\zeta_2)\pm P_+(\zeta_1,\zeta_2),
\end{equation}
the boundary q-KZ equations are simply expressed by
\begin{eqnarray}
  F_\epsilon(q^{-2}\zeta_1,\zeta_2) &=&
  \frac{\kappa(\zeta_1^2;-1)\kappa(q^{-2}\zeta_1^2;-1)}
  {\kappa(\zeta_1^{-2};-1)\kappa(q^2\zeta_1^{-2};-1)}
  f(\epsilon\zeta_1\zeta_2)f(\epsilon\zeta_1/\zeta_2)
  F_\epsilon(\zeta_1,\zeta_2),
  \\
  F_\epsilon(\zeta_1,q^{-2}\zeta_2) &=&
  \frac{\kappa(\zeta_2^2;-1)\kappa(q^{-2}\zeta_2^2;-1)}
  {\kappa(\zeta_2^{-2};-1)\kappa(q^2\zeta_2^{-2};-1)}
  f(\epsilon q^{-2}\zeta_2/\zeta_1)f(\epsilon\zeta_1\zeta_2)
  F_\epsilon(\zeta_1,\zeta_2),
\end{eqnarray}
with
\begin{equation}
  f(\zeta) =
  \frac{(-q\zeta^{-1};p)_\infty(-pq^{-1}\zeta;p)_\infty}
  {(-q\zeta;p)_\infty(-pq^{-1}\zeta^{-1};p)_\infty}g(\zeta^2).
\end{equation}
These difference equations are exactly solved \cite{JKKKM2}
and $F_\epsilon(\zeta_1,\zeta_2)$ is given by
\begin{equation}
  F_\epsilon(\zeta_1,\zeta_2) =
  X(\zeta_1)X(\zeta_2)Y(\epsilon\zeta_1\zeta_2)Y(\epsilon\zeta_1/\zeta_2),
\end{equation}
where
\begin{eqnarray}
  X(\zeta) &=&
  \frac{(q^2\zeta^{-2};p,q^4)_\infty(q^6\zeta^{-4};p,q^4,q^8)_\infty
    (pq^2\zeta^{-4};p,q^4,q^8)_\infty}
  {(p\zeta^{-2};p,q^4)_\infty(q^4\zeta^{-4};p,q^4,q^8)_\infty
    (pq^4\zeta^{-4};p,q^4,q^8)_\infty}
  \nonumber \\ && \times
  \frac{(q^4\zeta^2;p,q^4)_\infty(q^{10}\zeta^4;p,q^4,q^8)_\infty
    (pq^6\zeta^4;p,q^4,q^8)_\infty}
  {(pq^2\zeta^2;p,q^4)_\infty(q^8\zeta^4;p,q^4,q^8)_\infty
    (pq^8\zeta^4;p,q^4,q^8)_\infty},
  \\
  Y(\zeta) &=&
  \frac{(-pq^{-1}\zeta^{-1};p,q^2)_\infty(q^2\zeta^{-2};p,q^4,q^4)_\infty
    (pq^2\zeta^{-2};p,q^4,q^4)_\infty}
  {(-q\zeta^{-1};p,q^2)_\infty(q^4\zeta^{-2};p,q^4,q^4)_\infty
    (p\zeta^{-2};p,q^4,q^4)_\infty}
  \nonumber \\ && \times
  \frac{(-pq\zeta;p,q^2)_\infty(q^6\zeta^2;p,q^4,q^4)_\infty
    (pq^6\zeta^2;p,q^4,q^4)_\infty}
  {(-q^3\zeta;p,q^2)_\infty(q^8\zeta^2;p,q^4,q^4)_\infty
    (pq^4\zeta^2;p,q^4,q^4)_\infty}.
\end{eqnarray}

Substituting $-q^{-1}$ and 1 into $\zeta_1$ and $\zeta_2$ respectively,
we find the spontaneous magnetization in the half-infinite chain,
\begin{equation}
  \label{hm}
  \langle\sigma^z\rangle_\infty =
  -\frac{F_-(-q^{-1},1)}{F_+(-q^{-1},1)} =
  -\frac{(q^2;q^2)_\infty^4(-p;p)_\infty^4}
  {(-q^2;q^2)_\infty^4(p;p)_\infty^4}.
\end{equation}
Applying the finite size formula (\ref{dual})
we obtain the spontaneous magnetization in the bounded XYZ model:
\begin{equation}
  \label{result}
  \langle\sigma^z\rangle =
  -\frac{2(q^4;q^4)_\infty^4(p^2;p^2)_\infty^4}
  {(q^2;q^2)_\infty^8(-p;p)_\infty^8+(-q^2;q^2)_\infty^8(p;p)_\infty^8}.
\end{equation}
In the limit $p\rightarrow 0$ this degenerates into the result
in the XXZ model \cite{Fujii2}.
The magnetization (\ref{result}) is roughly twice
as large as that in the half-infinite chain (\ref{hm}),
when the anisotropies $\Gamma$ and $\Delta$ are
in the neighborhood of $0$ and $-1$, respectively.
We conclude that the finiteness of the model
enhances the magnetization.

\setcounter{equation}{0}
\section{Conclusion}
We have shown that the vertex operator approach is valid
for the bounded XYZ spin chain.
Further, considering the turning symmetry,
we have proved the finite size formula (\ref{dual}).
By the virtue of the finite size formula
the magnetization in the bounded chain
can be derived from that in the half-infinite chain.
We have thus obtained the spontaneous magnetization (\ref{result})
in the XYZ model with boundaries.

In \cite{Fujii2} it is shown that
the excited states are composed by use of the VOs of another type (VOs II).
To investigate the energy spectrum of excited states in the XYZ model
one needs the explicit expression of the VO II\@.
Recently, an elliptic algebra $U_{q,p}(\widehat{sl}_2)$ for
the elliptic RSOS model has been bosonized,
and the VOs I and II have been given in the free field expressions
\cite{K,JKOS}. (The VO appeared in this paper is the VO I.)
Referring to this result may enable us to deal with the VO II
of $\mathcal{A}_{q,p}(\widehat{sl}_2)$ directly.

We have interpreted the boost operator of the boundary q-KZ equation
as the transfer matrix and
have expressed the eigenstate in terms of VOs.
Applying the Baxter's formula (\ref{bf})
to such expression of the eigenstate
we have obtained its asymptotic form (\ref{sz}).
In return,
this means that we have indeed given the asymptotic form of solution
of the q-KZ equation.
In particular we have found
the asymptotic form of the Smirnov's form factor,
which was developed in the integrable quantum field theory \cite{S}.
We must make clear a new idea for the q-KZ equation,
which corresponds to the turning symmetry in the bounded model.

Up to the present,
only the one-dimensional Heisenberg spin chains have been considered.
The models in higher-dimensions do not satisfy
the Yang-Baxter equation, which ensures solvability.
However, in our method,
one can take the VO for a site of spin chain itself,
likewise the R-matrix in the vertex model.
Therefore we expect that
\textit{the two-dimensional products of the VOs
describe the eigenstates in the two-dimensional Heisenberg spin lattice}.
There, the Yang-Baxter equation does not play a central role
for finding the energy spectrum.
The commutation relation between VO I and VO II
determines the energy spectrum
and may give self-consistency equations similar to the Bethe equation.
By using the VOs and
also considering geometric symmetries besides quantum groups, 
new two-dimensional models may be analyzed.

\section*{Acknowledgment}

The authors thank to H. Ujino, M. Shiroishi, Y. Kajinaga and T. Tsuchida
for discussions.


\begin{thebibliography}{99}
\bibitem{CBMS} M. Jimbo and T. Miwa:
  \textit{Algebraic Analysis of Solvable Lattice Models}
  (CBMS Regional Conference Series in Mathematics \textbf{85}, AMS, 1994).
\bibitem{DFJMN} B. Davies, O. Foda, M. Jimbo, T. Miwa and A. Nakayashiki:
  \textit{Comm. Math. Phys.} \textbf{151} (1993) 89.
\bibitem{FIJKMY1} O. Foda, K. Iohara, M. Jimbo, R. Kedem, T. Miwa and H. Yan:
  \textit{Lett. Math. Phys.} \textbf{32} (1994) 259.
\bibitem{FIJKMY2} O. Foda, K. Iohara, M. Jimbo, R. Kedem, T. Miwa and H. Yan:
  \textit{Prog. Theor. Phys. Suppl.} \textbf{118} (1995) 1.
\bibitem{JMN} M. Jimbo, T. Miwa and A. Nakayashiki:
  \textit{J. Phys.} A \textbf{26} (1993) 2119.
\bibitem{LP} M. Lashkevich and Y. Pugai:
  \textit{Nucl. Phys.} B \textbf{516} (1998) 623.
\bibitem{JKKKM1} M. Jimbo, R. Kedem, T. Kojima, H. Konno and T. Miwa:
  \textit{Nucl. Phys.} B \textbf{441} (1995) 437.
\bibitem{JKKKM2} M. Jimbo, R. Kedem, T. Kojima, H. Konno and T. Miwa:
  \textit{Nucl. Phys.} B \textbf{448} (1995) 429.
\bibitem{FR} I. B. Frenkel and N. Reshetikhin:
  \textit{Comm. Math. Phys.} \textbf{146} (1992) 1.
\bibitem{S} F. A. Smirnov:
  \textit{Form Factors in Completely Integrable Models
    of Quantum Field Theory}
  (World Scientific, Singapore, 1992).
\bibitem{Fujii1} Y. Fujii and M. Wadati:
  \textit{preprint} \texttt{solv-int/9801026},
  to appear in \textit{Chaos, Solitons and Fractals}.
\bibitem{Fujii2} Y. Fujii and M. Wadati:
  \textit{preprint} \texttt{hep-th/9807201}.
\bibitem{Sk} E. K. Sklyanin:
  \textit{J. Phys.} A \textbf{21} (1988) 2375.
\bibitem{B} R. J. Baxter:
  \textit{Exactly Solved Models in Statistical Mechanics}
  (Academic Press, London, 1982).
\bibitem{IK} T. Inami and H. Konno:
  \textit{J. Phys.} A \textbf{27} (1994) L913.
\bibitem{K} H. Konno:
  \textit{preprint} \texttt{q-alg/9709013}.
\bibitem{JKOS} M. Jimbo, H. Konno, S. Odake and J. Shiraishi:
  \textit{preprint} \texttt{math.QA/9802002},
  to appear in \textit{Comm. Math. Phys.}.
\end{thebibliography}
\end{document}